\documentclass[reprint,superscriptaddress,amsmath,amssymb,aps,prl]{revtex4-1}

\usepackage{graphicx}
\usepackage{dcolumn}
\usepackage{bm}
\usepackage{hyperref}

\usepackage[dvipsnames]{xcolor}

\usepackage{xr}
\begin{document}
\title{Rheology and Thermal Response of Entangled DNA Origami Tadpoles}
\title{Rheology and Programmable Gelation of DNA Origami Polymer Tadpoles}

\author{Jennifer Harnett}
\affiliation{School of Physics and Astronomy, University of Edinburgh, Peter Guthrie Tait Road, Edinburgh, EH9 3FD, UK}
\author{Saminathan Ramakrishnan}
\affiliation{School of Physics and Astronomy, University of Edinburgh, Peter Guthrie Tait Road, Edinburgh, EH9 3FD, UK}
\author{Alice L. B. Pyne}
\affiliation{School of Chemical, Materials and Biological Engineering, University of Sheffield, Sheffield, UK.}
\author{Elizabeth P. Holmes}
\affiliation{School of Chemical, Materials and Biological Engineering, University of Sheffield, Sheffield, UK.}
\author{Davide Michieletto}
\affiliation{School of Physics and Astronomy, University of Edinburgh, Peter Guthrie Tait Road, Edinburgh, EH9 3FD, UK}
\affiliation{MRC Human Genetics Unit, Institute of Genetics and Cancer, University of Edinburgh, Edinburgh EH4 2XU, UK}
\affiliation{International Institute for Sustainability with Knotted Chiral Meta Matter (WPI-SKCM$^2$), Hiroshima University, Higashi-Hiroshima, Hiroshima 739-8526, Japan}

\begin{abstract}
DNA origami is a powerful method to achieve nanoscale folded structures. Despite rapid improvements in folding and purification methods, DNA origami objects are still often produced in small quantities and studied at single molecule scale.
Here, we design simple DNA origami-inspired polymers with complex topologies, and study their rheology and viscoelastic properties in dense conditions. 
First, we designed and purified topologically distinct DNA nanostructures -- linear, circular, and ``tadpole'' polymers -- to evaluate how polymer architecture influences entanglement and rheology. Despite their distinct topologies, we observe that all constructs obeyed universal rheological scalings, likely due to their short length. However, upon thermal annealing in the bulk, the DNA origami-like polymers displayed significantly different behaviours. Our results suggest that DNA origami-like polymers could be used to engineer thermoresponsive behaviours in complex fluids by introducing reversible and topology-dependent crosslinking. 
\end{abstract}

\maketitle

\section*{Introduction}

Despite decades of research to develop the synthesis of polymers with non-trivial topologies and architectures,  to date the vast majority of polymeric materials rely on simple linear polymers~\cite{Schroeder2026RingPolymerRheology}.  This is mostly due to three main challenges: first,  synthetic pathways to create precisely controlled polymers with complex architectures are often cumbersome and require extensive experience and know-how~\cite{Doi2015tadpole,Doi2020tadpole}; second,  purification and isolation of architecturally complex polymers from linear ones is not straightforward~\cite{Kapnistos2008}; third,  the presence of even small contaminants of linear chains is often enough to dominate the rheology of the solution~\cite{Kapnistos2008}. 

Unconventional and architecturally complex polymers include ring polymers, i.e. polymers whose free ends are closed in a loop~\cite{Kapnistos2008,Goossen2014} and so-called ``topological'' polymers, which are formed by gluing together ring and linear polymers~\cite{Deguchi2017}. Both these families contain architectures with looped sub-structures, where part of the polymer contour is topologically closed into a loop. The physics of ring polymers, i.e. of polymers containing no free ends,  has challenged the polymer physics community for decades~\cite{Cates1986,Smrek2015a,Halverson2011c,Halverson2011dynamics,Halverson2014,Ge2016}. Due to the lack of free ends, they do not follow the well-established reptation dynamics in the melt~\cite{Gennes1979,Doi1988b}. Additionally, because of their closed topology, they can display unconventional topological constraints called threadings~\cite{Michieletto2014,Tsalikis2016SlowModesRingPolymers}. Threadings,  unlike more conventional entanglements in linear polymers, can induce dynamical heterogeneities and glass-like behaviours at temperatures much higher than the glass transition temperature of the system~\cite{Michieletto2016pnas,Michieletto2017prl}.  Beyond theoretical curiosity,  solutions of ring polymers display unconventional rheological behaviours in simulations, for example, ultra softness when crosslinked into an elastomer~\cite{Wang2022}, extreme non-linear shear thickening in extensional flow~\cite{OConnor2020}, and also affect the phase separation kinetics and thermodynamics of diblock co-polymers~\cite{Wijesekera2024}. 

In blends of ring and linear polymers,  linear chains dominate the relaxation time of the system~\cite{Kapnistos2008, OConnor2022} and mild cooperative behaviour is found in the regime of small ring fraction, in which case the solution displays a moderate two-fold thickening with respect to a solution of pure linear chains~\cite{Parisi2020}.

Whilst these results may suggest that a system containing both linear and ring polymers may not offer interesting rheology,  it is interesting to consider polymer designs where the ring and linear architectures are fused together in ``topological'' (or chimeric) polymers~\cite{Rosa2020}.  Dense solutions of the simplest ``topological'' polymer, i.e. a tadpole-like structure has been studied \emph{in silico}~\cite{Rosa2020} and in experiments~\cite{Doi2015tadpole,Doi2020tadpole}. Both computer simulations and experiments suggest that threadings are abundant in dense solutions of tadpoles, and that they trigger a cooperative slowing down due to the formation of hierarchical constraints which are unique to this polymer architecture~\cite{Rosa2020,Doi2015tadpole,Doi2020tadpole}. 
However, the synthesis of pure solutions of tadpole-like synthetic polymers remains a very difficult task and higher order topological polymer designs would be even more challenging to realise and purify. 

In an attempt to address the open challenge to synthesise scalable quantities of pure topological polymers with well defined and complex architectures,  we here explore the use of DNA origami.  Specifically, we focus on the simplest topology, i.e. a DNA origami tadpole  (see Fig.\ref{fig:origamidesign}). We optimised the yield of these DNA-origami polymers and in turn managed to prepare solutions tadpole polymers at $10 C^*$, i.e. 10 times the overlap concentration of the polymers and thus well within in the entangled regime.  

To the best of our knowledge,  this is the first time that DNA origami has been considered to create polymers with complex topologies and to study their rheology at high concentrations.  As we show in this paper,  we do not observe topology-dependent rheological behaviours; however, we argue that this result is due to the insufficient length of the chosen DNA origami scaffold.  At the same time,  by performing thermal annealing and quenching we do observe topology-dependent gelation pathways that can be designed and programmed through the choice of DNA staples.
 
Overall, we argue that our approach offers a new way of realising polymers with complex architectures by leveraging Watson-Crick base-pairing and thus bypassing the need of inventing new synthesis pathways.  We also argue that in the future it will be possible to make DNA origami designs using multiple scaffolds so that they can be merged into higher-order,  larger origami polymers.

\section{Materials and Methods}
\subsection{DNA Origami Design and Preparation}
\begin{figure*}
\centering
{\includegraphics[width=0.9\textwidth]{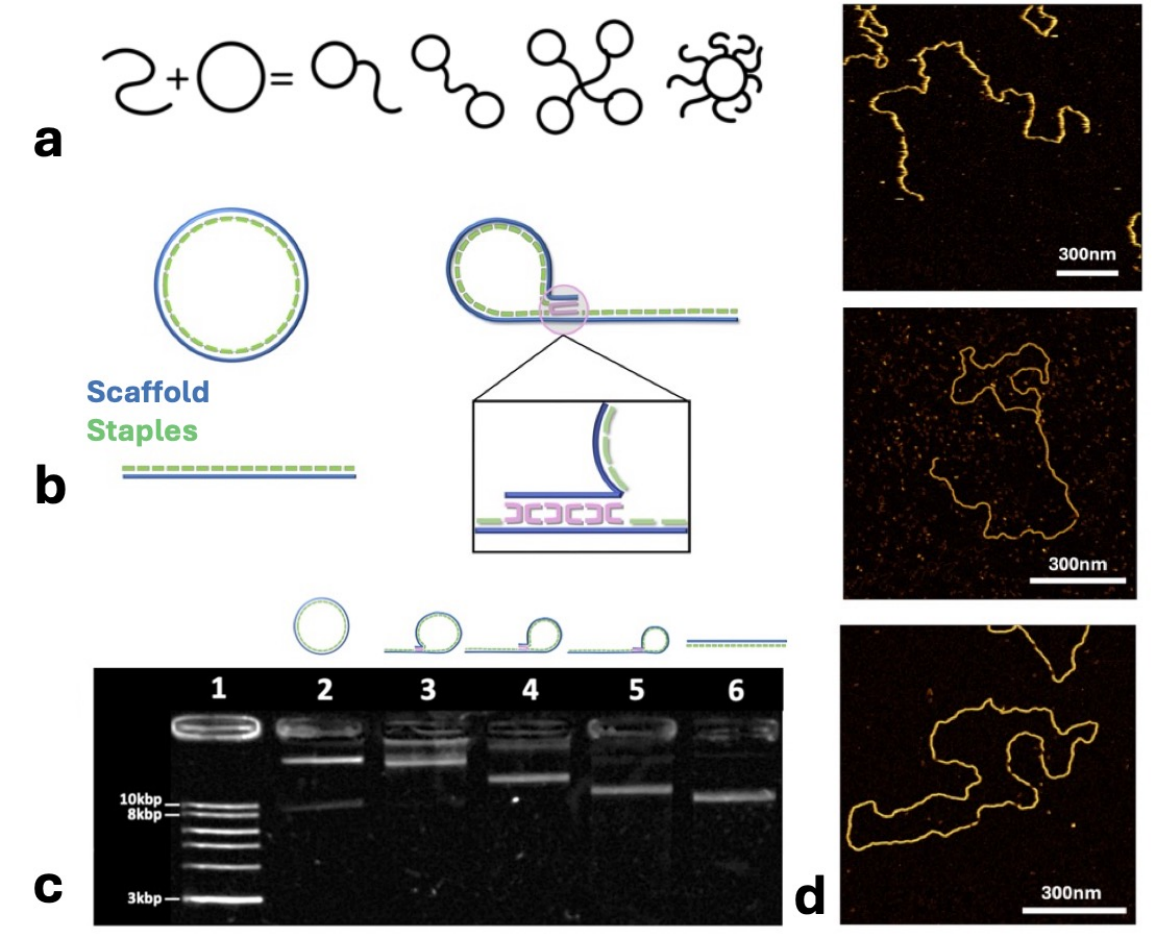}}
\caption{\textbf{DNA origami for the synthesis of topological polymers.} \textbf{(a)} Examples of topological polymers made by the gluing of circular and linear architectures.  \textbf{(b)} Design of three distinct DNA origami topologies: circular, linear, and tadpole. The single stranded DNA (ssDNA) M13m18 scaffold (8064 bp) is shown in blue, and the complementary ssDNA staples are shown in green.  The circular and linear topologies are prepared using the same DNA staples and differ only on the topology of the scaffold.  The tadpole topology has 189 staples tiling the head and tail sections, and 6 staples (pink) to stabilise the loop.  The number of staples depicted is not representative of the true number of staples, which is $\approx$ 200 per origami structure. \textbf{(c)} Gel electrophoresis confirms the folding of our designs.  From left to right: 1kbp molecular weight ladder control (1), circular origami (2),  $\frac{3}{4}$ tadpole (3), $\frac{1}{2}$ tadpole (4), $\frac{1}{4}$ tadpole (5), linear (6). (d) AFM in liquid of linear, tadpole, and circular DNA origami topologies (from top to bottom).}
\label{fig:origamidesign}
\end{figure*}

The DNA origami structures were designed using cadnano software (cadnano.org) with the help of Tilibit Nanosystems who also provided the scaffold and staples.  Figure~\ref{fig:origamidesign}b shows simplified design of each topology: circular, linear and tadpole (where 50\% of the scaffold length comprises the tadpole head and 50\% the tail section). Both the linear and circular designs were prepared using the same staple stock, containing 193 staples (ranging in length from 36-48 bases). The tadpole staple stock contained 189 staples (36-48 bases) and an additional 6 shorter staples (23-26 bases) in the crosslinking section to loop the scaffold on itself.

All DNA origami samples were prepared using a 8064-base long single stranded DNA (ssDNA) phage vector (M13mp18) scaffold widely used in DNA origami protocols~\cite{wagenbauer2017we}. To prepare the linear and tadpole topologies we first cut the circular scaffold using a single staple that creates an EcoRI recognition site; upon cleavage with EcoRI-HF, we obtain a linear scaffold. 
Specifically, a 23-base ssDNA oligomer, with the sequence 5' AACTAGAGCTCGAATTCGTAATC 3', was annealed using the following temperature ramp: 65$^\circ$ C  for 15 minutes, 64$^\circ$ C – 25$^\circ$ C  (-1$^\circ$ C  / 1 min). Reaction concentrations were: 85 nM scaffold, 680 nM oligomer, 1X CutSmart Buffer (50 mM Potassium Acetate, 20 mM Tris-acetate, 10 mM Magnesium Acetate, 100 $\mu$g/ml BSA (Bovine Serum Albumin) ) (NEB B7204). Next, the circular scaffold was digested into a linear scaffold. The cutting mixture contained: 50 nM scaffold, 0.4 $U/\mu$L EcoRI-HF, and 1X rCutSmart buffer (50 mM Potassium Acetate, 20 mM Tris-acetate, 10 mM Magnesium Acetate, 100 $\mu$g/ml Recombinant Albumin) (NEB B6004). This mixture was incubated at 37$^\circ$ C for 60 minutes, followed by a 20 minute inactivation step at 65$^\circ$ C . After the digestion step, ethanol precipitation was performed to concentrate the DNA, and the DNA pellet was resuspended in 1X TE buffer (10 mM Tris-HCl, pH 8, 1 mM EDTA). The concentration of the ssDNA scaffold was determined using a Nanodrop-Lite UV spectrophotometer at 260 nm. 

Since the ratio between the size of the tadpole head and tail can drastically affect the presence of entanglements in dense solutions~\cite{Rosa2020}, we also designed two additional tadpole topologies: one with the tadpole head comprising 25\% of the scaffold length ($\frac{1}{4}$ tadpole), and one with the tadpole head comprising 75\% of the scaffold length ($\frac{3}{4}$ tadpole). 

A total folding reaction of 100 $\mu$L was prepared in 200 $\mu$L PCR tubes. A folding buffer (named 1X FoB20, as in \cite{wagenbauer2017we}) was added (1 mM EDTA, 5 mM Tris Base, 5 mM NaCl, 20 mM MgCl$_2$, pH 8) alongside 20 nM scaffold and 200 nM staples (both supplied by Tilibit Nanosystems). The origami was annealed using a temperature ramp: 65$^\circ$ C for 15 minutes, and 60 $-$ 44$^\circ$ C (at a rate of -1$^\circ$ C per hour).  Samples were stored at -20$^\circ$ C. The folding was then checked on a 1.5\% agarose gel (see Fig.~\ref{fig:origamidesign} and next section). Concentrated DNA origami was stored in a buffer (named 1X FoB5) containing 1 mM EDTA, 5 mM Tris Base, 5 mM NaCl and 5 mM MgCl$_2$ for stability and to prevent aggregation \cite{wagenbauer2017we}. 

\subsection{Gel Electrophoresis for Analysis of DNA Topologies and Mass Quantification}

Gel electrophoresis was performed using a 1X TAE buffer (40 mM Tris, 20 mM acetic acid, and 1 mM EDTA) with 5 mM MgCl$_2$ added for DNA origami stability~\cite{wagenbauer2017we}. Without the presence of magnesium, all tadpole DNA origami topologies would run in line with the linear topology. A 1.5\% (w/v) agarose gel was required to resolve the different topologies. A 6X solution of loading buffer was prepared containing: 0.25\% Bromophenol Blue and 15\% Ficoll-400. 50 ng of the sample DNA was loaded into each well and a 1 kbp ladder (NEB N3232S) was used as the molecular weight marker control. The gel was run at 4 V/cm for 3 hrs. Following electrophoresis, the gel was immersed  in a solution of 1X TAE buffer with a 1:10,000 dilution of SYBR Gold Nucleic Acid Gel Stain (10,000X Concentrate in DMSO) (Thermofisher S11494), and placed on a shaker for 1 hr. Images were taken using a UV transilluminator with an exposure of 100 ms. 

The DNA origami samples contain excess ssDNA staples in solution, and therefore it was not possible to accurately determine the mass of DNA origami using UV spectrophotometry. There are various methods established for the purification of DNA origami~\cite{wagenbauer2017we}, however, most result in a loss of DNA origami mass. It was essential to maximise the yield of DNA origami to obtain enough material to create entangled solutions for rheological studies.  Therefore, we decided to skip the purification of excess staples as, arguably,  they should not affect the architecture and therefore the rheology once the samples have been quenched at room temperature. 

To estimate the yield and therefore the mass of the DNA origami topologies, we used gel electrophoresis and compared the intensity of the tadpole bands to that of a standard molecular weight ladder.  ImageJ was used to calculate an intensity-to-mass conversion using six bands from the 1 kbp ladder of known mass, and an error determined from the standard deviation. At least three independently prepared agarose gels were imaged under identical conditions with a $\pm10\%$ error observed between the gels. 

\subsection{Atomic Force Microscopy}
To characterise the correct folding of the DNA origami tadpoles we performed atomic force microscopy (AFM) to image them at single molecule resolution. 

For imaging in air, the samples were diluted to 2 ng/$\mu$L in a buffer containing 10 mM MgCl$_2$. The DNA was adsorbed onto mica (5 mm diameter) for 3 minutes. The mica disk was then washed once using ultrapure water to remove any unbound structures, and dried using nitrogen air. ScanAyst-HR Air cantilevers were used with, k=0.4 N/m, f=130 kHz and 2 nm radius (where k is the cantilever spring constant, and f the resonance frequency). All AFM air images were taken either on a MultiMode 8-HR (Bruker) AFM at TU Delft, or JPK NanoWizard at the University of Edinburgh. 

For imaging in liquid, the sample was diluted to 2 ng/$\mu$L in a buffer containing 25 mM MgCl$_2$, 10 mM Tris, and 1 mM EDTA (pH 7.4). The sample was adsorbed onto the mica for 30 minutes, and then exchanged three times with a fresh imaging buffer (containing 3 mM NiCl$_2$ and 20 mM HEPES, pH 7.4). This method ensured that all topologies were absorbed to the mica in an open conformation. A FastScan-D cantilever probe was used with a 1 nm tip radius, high resonant frequency (110 kHz in fluid), and low spring constant (0.23 N/m). Peak force tapping mode was used with a force of 70-100 pN selected, so that the helix was not deformed \cite{pyne2014single}. Images in liquid were taken either using a FastScan Bio AFM system (Bruker), or a JPK NanoWizard. All AFM images were post processed using Gwyddion adapting methods from \cite{Gwyddion}.

\subsection{Microrheology}
To quantify the rheology of DNA origami tadpoles, we performed microrheology on $\approx$ 5 $\mu$L samples.  Given the small amounts of DNA origami that could be purchased and folded, it would not be possible to perform bulk rheology on mL samples. 

Serial dilutions were prepared by transferring the stock solution into 1X Fob5 buffer, to reach the final desired concentration. The range of concentration spans that of the dilute, semi-dilute unentangled, and semi-dilute entangled regime. To perform microrheology we spiked BSA-coated 1.1 $\mu$m diameter particles in the samples and loaded 3-5 $\mu$L of sample into a chamber created on a glass slide using double sided tape (100 $\mu$m thick). Videos were recorded of the Brownian motion of the particles embedded in fluids at different DNA concentration and polymer topology. Videos were taken on a Nikon Eclipse TS2 microscope with a 60X objective. All microrheology experiments were performed in a temperature-controlled Okolab chamber stage-top incubator at 25$^{\circ}$C.

We discovered that 400 fps (frames per second) videos were necessary to observe $G'>G''$ in entangled DNA origami. However, recording long 400 fps videos was not practical due to the extremely large file size, therefore, short-time behaviour was captured by recording a 400 fps video for 10 s, and long-time behaviour using a 2 fps video for up to 5 minutes.  

The particles were then tracked using Trackpy library (github.com/soft-matter/trackpy) and custom-written C++ particle-tracking codes. The time averaged mean squared displacement (MSD) of the particles was measured as a function of lag time $t$ as
\begin{equation}
MSD(t)  = \langle  \left[\bm{r}_i(t + t_0) - \bm{r}_i(t_0) \right]^2 \rangle
\end{equation} 
where the average is performed over time $t_0$ and particles $i$. 
We then averaged the MSD along $x$ and $y$ directions,  and used it to extract the diffusion coefficient $D$ of the beads as $D = \lim_{t \to \infty} MSD(t)/2t$. The MSD was fitted at a large lag time and computed using 3 different lag time ranges and then averaged. The Stokes-Einstein equation $\eta = k_B T/(3 \pi D a)$ was used to estimate the viscosity of each sample. 

The viscous and elastic moduli were calculated by using the generalised Stokes-Einstein relation~\cite{mason2000estimating}. Briefly, we fitted the MSD using a polynomial function and we then extracted the complex modulus as
\begin{equation}
    |G^{*}(w)| = \dfrac{k_B T}{3 \pi a MSD(1/w) \Gamma[ 1 + \alpha(w)]}
\end{equation}
where $\alpha(w) = \left. d \log{MSD(t)} / d \log{t} \right|_{t = 1/w}$ is the MSD exponent as a function of lagtime and $\Gamma$ is the Gamma function. The viscous $G^{''}$ and elastic $G^{'}$ moduli are then computed as 
\begin{align}
    & G^{'}(w) = |G^{*}(w)| \cos{(\pi \alpha(w)/2)} \\
    & G^{''}(w) = |G^{*}(w)| \sin{(\pi \alpha(w)/2)} \, .
\end{align}

The 400 fps videos were used to determine $G^{'}$ and $G^{''}$, and the 2 fps videos used to estimate a value for $\eta$. 

\section{Results}

\subsection{Creating Unique DNA Topologies with DNA Origami}
First, we folded and characterised the topology of the designed DNA origami polymers. As described in the Methods,  we followed well established DNA origami protocols to fold our structures~\cite{wagenbauer2017we}.  By using gel electrophoresis, we observed that the linear, circular, tadpole (1/4, 1/2 and 3/4) could all be separated because they run at different speed along the gel (Fig.~\ref{fig:origamidesign}c). This is due to the size of their circular component~\cite{michieletto2015rings}. The circular DNA origami sample was estimated to contain $\approx$ 25\% linear contamination.

The circular and linear DNA origami were concentrated to concentrations $\approx$ 10$C^*$ using ethanol precipitation. Ethanol precipitation was too harsh to conserve the tadpole topology, and resulted in the structure becoming aggregated, therefore, the tadpole topology was concentrated using Amicon 50K filters. Detailed methods on the concentration of DNA origami topologies can be found in the Supplementary Information. 

Finally, to further characterise the structures we also performed AFM, confirming that our structures were folded in the way we designed them to be (fig.~\ref{fig:origamidesign}d). 

\subsection{The solutions' rheology is topology independent}

Based on recent computational~\cite{Rosa2020} and experimental work~\cite{Doi2015tadpole,Doi2020tadpole}, we expected dense solutions of tadpoles to display significantly different dynamics with respect to their linear and circular counterpart. Specifically, we hypothesised that the tadpoles solutions would display a slower dynamics (and hence larger viscosity) than either linear and ring solutions~\cite{Rosa2020,Doi2020tadpole}.  

\begin{figure*}
{\includegraphics[width=0.9\textwidth]{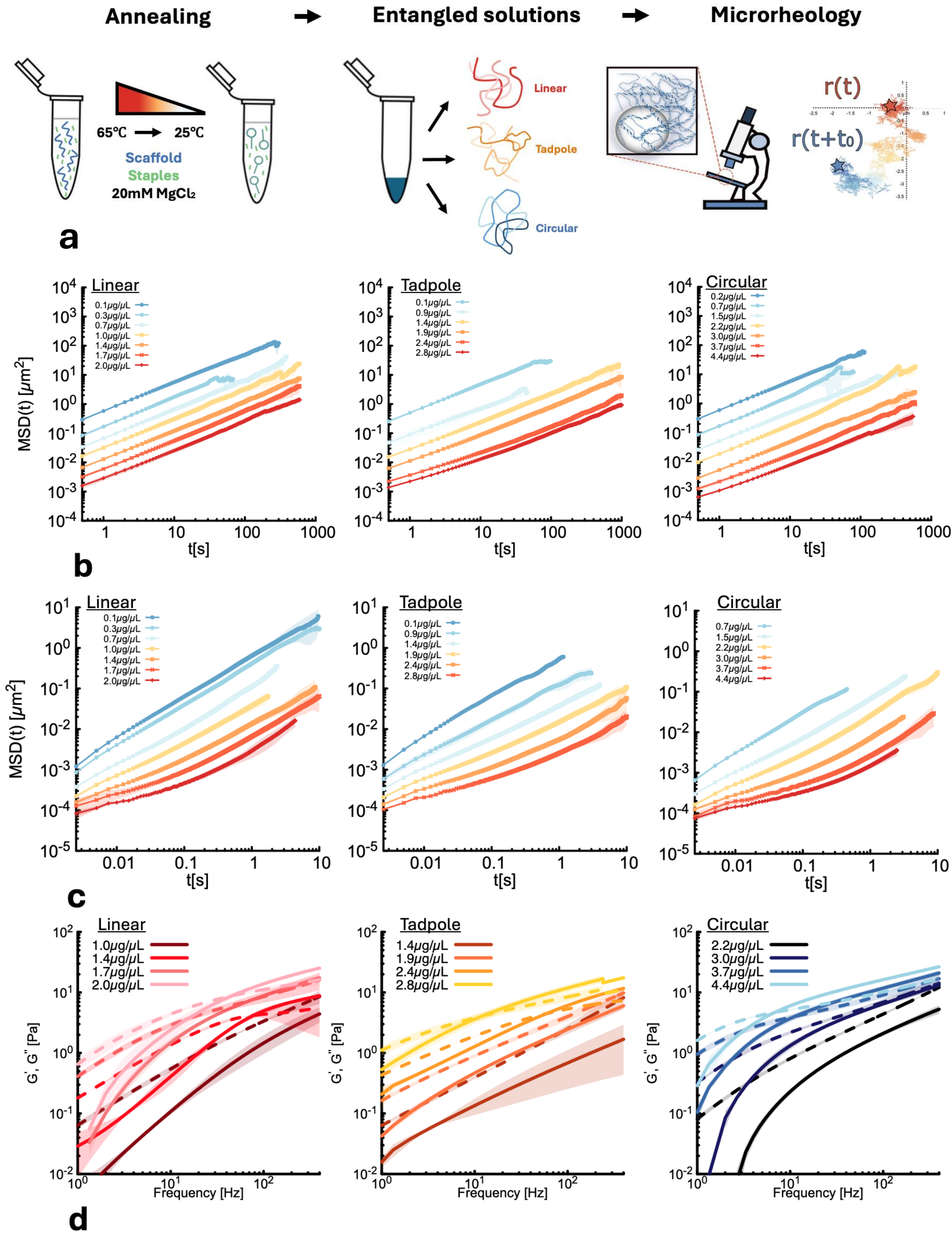}}
\caption{\textbf{Microrheology suggests topology-independent rheology.} (a) Sketch of experimental workflow. (b) Mean squared displacement (MSD) as a function of lag time recorded at 2 fps for DNA origami of varying topologies: linear, tadpole and circular topologies (left to right). (c) MSD as a function of lag time recorded at 400 fps of linear, tadpole and circular topologies (left to right). The MSD for increasing DNA concentration is shown for each topology. (d) Storage modulus ($G'$) shown in solid lines and loss ($G''$) moduli shown in dashed lines, as a function of frequency (Hz). Data for linear, tadpole, and circular topologies is shown at varying DNA concentrations. For all samples the standard deviation with respect to positions in the same sample and across at least two samples repeats is represented on the plot by a shaded region, however for most samples this error was too low to be visualised.}
\label{fig:MSD_TLC_2fps}
\end{figure*}

To test this hypothesis and quantify the solutions' viscoelasticity for different topologies and DNA concentration, we performed microrheology and recorded both short 400 fps video and longer 2 fps videos to monitor the behaviour of the beads at different timescales. The MSD of the beads are reported in Fig.~\ref{fig:MSD_TLC_2fps}, where we show tracking at 2 fps (Fig.~\ref{fig:MSD_TLC_2fps}b) and 400 fps (Fig.~\ref{fig:MSD_TLC_2fps}c) and and across different topologies.  Finally, in Figure~\ref{fig:MSD_TLC_2fps}d, we show the behaviour of selected $G^\prime$ and $G^{\prime \prime}$ curves, obtained as described in the methods~\cite{mason2000estimating}.

\begin{figure*}
{\includegraphics[width=0.8\textwidth]{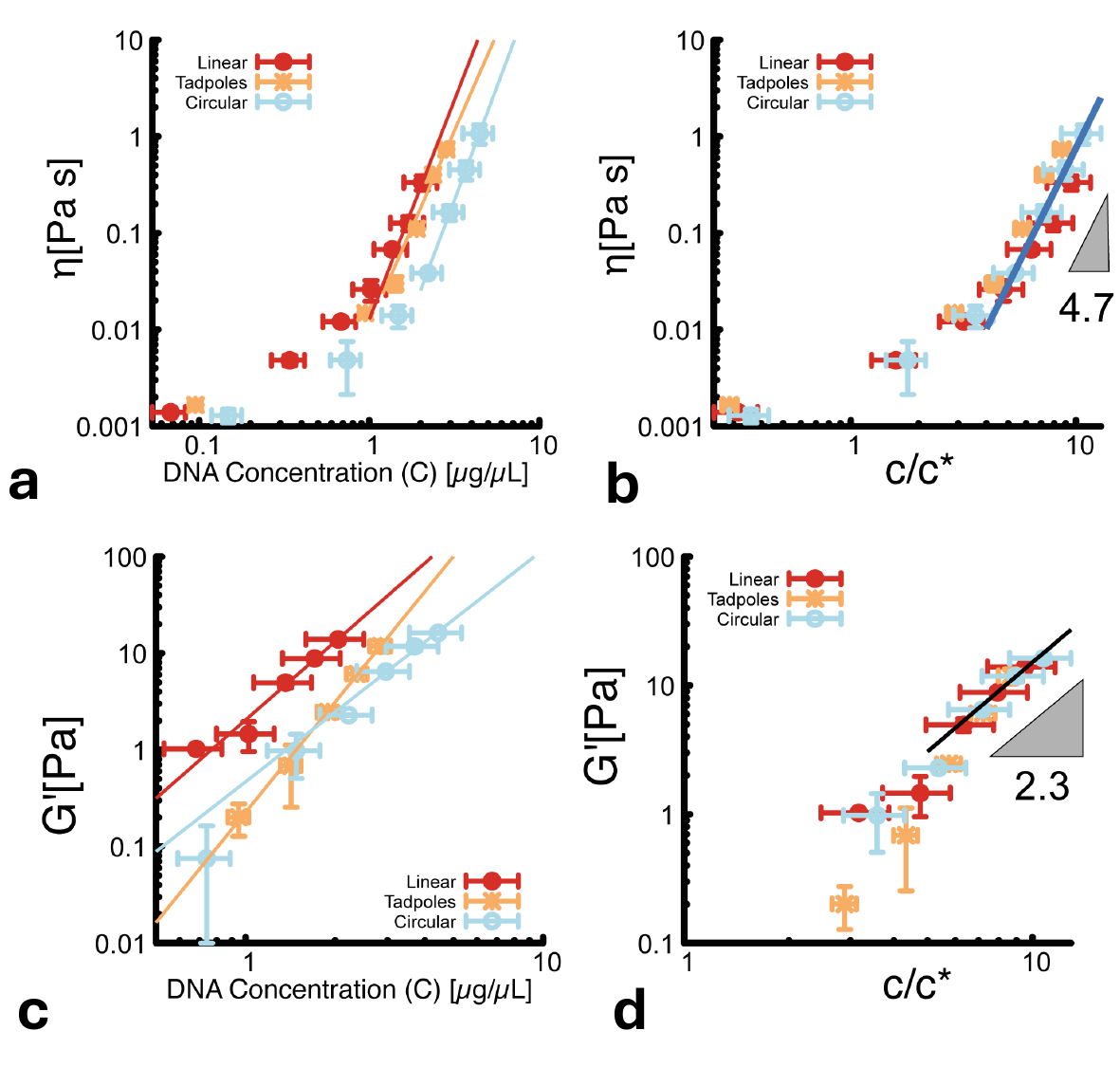}}
\caption{(a) Scaling of solution viscosity ($\eta$) with increasing concentration (C) of DNA for linear, tadpole and circular topologies. The data was fitted to the function $\eta(C)=AC^b$ for each topology for concentrations $C>4C^*$. We find $\eta(C)_{LINEAR} \sim C^{4.53 \pm 0.57}$, $\eta(C)_{TADPOLE} \sim C^{4.02 \pm 0.48}$ and $\eta(C)_{CIRCULAR} \sim C^{4.69 \pm 0.04}$. \textbf{(b)} The same data as in (a) but the DNA concentration is normalised with respects to a topology dependent $C^*$. The predicted scaling above $C_e$ is $\eta \sim C^{4.7}$ (blue line). (c) Value of Storage moduli ($G'$) at a 100 Hz with increasing DNA concentration for linear, tadpole and circular topologies. The data for each topology is fitted using the equation $G'(C)=AC^b$. We find $G'(C)_{LINEAR} \sim C^{2.69 \pm 0.19}$, $G'(C)_{TADPOLE} \sim C^{3.79 \pm 0.10}$ and $G'(C)_{CIRCULAR} \sim C^{2.40 \pm 0.47}$. (d) Same data as (c) with $C$ normalised via the overlap concentration ($C^*$) for the specific topology. The predicted scaling for synthetic entangled linear polymers is $G_p' \sim C^{2.3}$ (black line).}
\label{fig:master}
\end{figure*}

From the large time behaviour of the MSDs we obtained the zero shear viscosity of the solution through the Stokes-Einstein relation (see Methods), while the elastic plateau is estimated as the value of $G^\prime$ at the largest frequency monitored (i.e. the inverse of shortest time, in practice we use 100 Hz). 

From polymer physics theories~\cite{rubinstein2003polymer}, we expect both the zero shear viscosity ($\eta$) and elastic plateau modulus ($G_p'$) of polymer solutions scale with the concentration ($C$). The scalings are expected to follow distinct regimes depending on whether the solutions are dilute, semi-dilute unentangled, or semi-dilute entangled~\cite{Doi1988b}.  The theoretical predictions for the scaling of viscosity with $C$ in the Rouse and reptation regimes are $\eta \sim C^{1.3}$ and $\eta \sim C^{3.9}$ respectively~\cite{Doi1988b,rubinstein2003polymer}.  

However, the reptation model is known to underestimate (and in fact neglect) chain-chain correlations, such as ``double reptation''~\cite{DesCloizeaux1988} and dynamical entanglement~\cite{Michieletto2021entanglement}, and in turn it underestimates scaling exponents~\cite{rubinstein2003polymer}. For example, Takahashi reported $\eta \sim C^{4.5}$ for entangled poly(a-methylstyrene) in good solvents~\cite{takahashi1985zero}. Raspaud presented data on the zero shear viscosity ($\eta_0$) of three different synthetic polymers (all linear topology) of varying molecular weight~\cite{raspaud1995number} and reported that the data collapsed on a single universal curve when the concentration was normalised by the system-specific entanglement concentration $C_e$. The relation between $\eta_0$ and polymer concentration was predicted to scale as~\cite{raspaud1995number}: 
\begin{equation}
    \eta_0/\eta_{rouse} \sim \left(\frac{C}{C_e}\right)^{3.4},  \, 
    \label{raspaud}
\end{equation}
with 
\begin{equation}
    \eta_{rouse}= \eta_s \left(\frac{C}{C^*}\right)^{1/3\nu -1}, \, 
    \label{etarouse}
\end{equation}
and where $C_e$ is the entanglement concentration and $\eta_{rouse}$ is the Rouse viscosity of an unentangled polymer solution of the same molecular weight. The value of $\eta_{rouse}$ is dependent on polymer concentration (eq.~\ref{etarouse}), and therefore $\eta_0$ is predicted to scale as $\eta_0 \sim C^{4.7}$ at concentrations above $C_e$ (when using a Flory exponent of $\nu =0.588$). This scaling is in good agreement with experimental data on entangled solutions of linear T2 DNA (164 kbp) ~\cite{musti1995viscoelastic}, and a scaling of $\eta_{sp} \sim C^{4.3}$ was discovered for calf-thymus DNA ~\cite{bravo2016conformation}. Finally, rheological studies on entangled solutions of $\lambda$-DNA have been performed up to concentrations as high as 90$C^*$~\cite{Banik2021,Harnett2024}, finding scalings $\eta_{sp} \sim C^{3.2}$ up to 3$C^*$ and $\eta_{sp} \sim C^{5.5}$ above 3$C^*$ ~\cite{Banik2021}. 

\begin{figure*}[t!]
	\centering
	\includegraphics[width=0.95\linewidth]{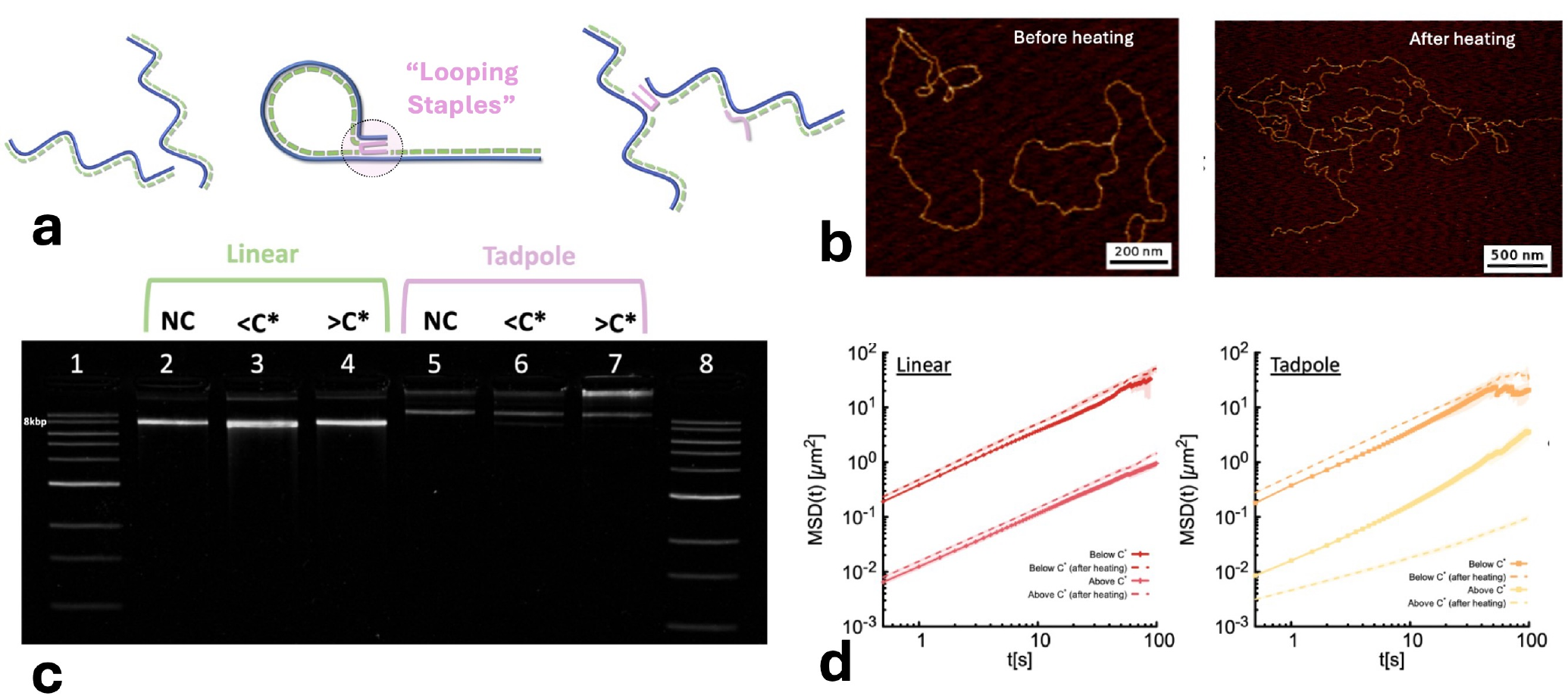}
	\caption{\textbf{Programmable gelation by \emph{in situ} thermal re-folding.} (a) Inter-molecular interactions between DNA origami scaffolds are mediated by ``bridging'' staples present in the tadpole design but absent in the linear and circular designs. (b) AFM of tadpole sample before and after thermal reconfiguration. (c) Gel electrophoresis of linear and tadpole topologies before and after thermal reconfiguration. The lanes are: 1 kbp ladder control with the location of 8 kbp labelled (1), linear origami before heating (negative control, NC) (2), linear origami after heating 70 $^\circ C$ at $C<C^*$ ($\approx 0.1 \ \mu g/\mu L$) (3), linear origami after heating 70 $^\circ C$ at $C>C^*$ ($\approx 1 \ \mu g/\mu L$) (4), tadpole origami NC before heating (5), tadpole origami after heating 70$^\circ C$ at $C<C^*$ ($\approx 0.1 \mu g/\mu L$) (6), tadpole origami after heating 70$^\circ C$ at $C>C^*$ ($\approx 1 \mu g/\mu L$) (7), 1 kbp ladder (8). (d) Comparison of linear and tadpole topology DNA origami before (solid lines) and after heating to 70$^{\circ}C$ and cooling to room temperature (dashed lines). Data is presented above $C^*$ ($\approx 1\ \mu g/\mu L$), and below $C^*$ ($\approx 0.1\ \mu g/\mu L$) for each topology. Videos taken at 2 fps. The shaded region corresponds to the standard deviation between repeats within the same sample and between at least two replicas.}
	\label{fig:gelation}
\end{figure*}

To compare our results with these theoretical and experimental observations, we gathered all the zero shear viscosity $\eta$ and elastic plateau $G^\prime$ data from the different topologies and concentrations (Fig.~\ref{fig:master}a,c) and re-plotted the data by normalising the concentration $C$ by the topology-specific overlap concentration $C^*$ (see section `Calculating a topology dependent $C^*$' in Supplementary Information).   
Figures \ref{fig:master}b,d show that the data points collapse (within error bars due to errors in estimating DNA concentration) on top of each other onto a master curve.  The data is in good agreement with the scaling predictions of $\eta \sim C^{1.3}$ in the semidilute unentangled regime (equation \ref{etarouse}) and $\eta \sim C^{4.7}$ in the semidilute entangled regime (equation \ref{raspaud}) (lines in Fig.~\ref{fig:master}). We observe a clear crossover from dilute to entanglement regime ($C_e$) at approximately 4$C^*$ which is in agreement with previous estimates for 25 kbp and 45 kbp DNA $C_e \approx 6C^*$~\cite{Robertson2007self} and for entangled $\lambda$-DNA $C_e \approx 3C^*$~\cite{zhou2018dynamically}. 

The elastic plateau modulus is predicted to scale as~\cite{rubinstein2003polymer}
\begin{equation}
G_p' \sim C^{3\nu/(3\nu -1)} \sim C^{2.3}
    \label{G'p}
\end{equation}
where the latter holds for linear polymers in good solvent ($\nu = 0.588$)~\cite{raspaud1995number}. This scaling has also been confirmed for synthetic linear polymers~\cite{raspaud1995number} as well as for entangled solutions DNA~\cite{mason1998linear, Banik2021}. Figure \ref{fig:master}b,d show the scaling of $G_p'$ for samples of varying topology and length of DNA normalised with respects to their value of $C^*$. We again see good agreement between the behaviour of the entangled DNA solutions and that reported for synthetic linear polymers and linear DNA~\cite{raspaud1995number,Banik2021}. 

While these results appear to be at odd with previous simulations~\cite{Rosa2020} and experiments~\cite{Doi2020tadpole} of tadpole polymers, we argue that our topology-insensitive observations are due to the modest length of our origami structures. Even at large concentrations of several $\mu$g/$\mu$l (the maximum we achieved in this work), the entanglement length of DNA solutions is of the order of 2 to 3 kbp (see Ref.~\cite{Fosado2022ihf}). Given that the M13 origami scaffold is about 8 kbp in total, this means that each polymer experiences at most 3 to 4 entanglements with neighbouring chains, which is insufficient to develop a deeply threaded state.

Furthermore, previous work investigating the diffusion of entangled DNA molecules \cite{robertson2007strong, Robertson2007self} concluded that 6 kbp and 11 kbp DNA molecules were too short to exhibit strong entanglement, showing Rouse-like behaviour up to 1 $\mu g/\mu L$, consistent with our results. However, we note that this study did not investigate concentrations higher than 1 $\mu g/\mu L$, above which in this study we find a crossover to an entangled reptation-like regime. This suggests that while the system is entangled, varying DNA topology does not introduce threading effects but instead only influences the effective molecular size ($R_g$), which determines the onset of entanglement.

\subsection{Programmable gelation by thermal in situ reconfiguration of DNA origami tadpoles}

Having observed that different topologies do not lead to distinct rheology in dense solutions, we then conjectured we could use the different DNA origami designs to create programmable gelation. Specifically, we realised that in the linear (and circular) DNA origami design, all the staples tile consecutive regions of the scaffold, whereas in the tadpole design, the overlap staples are complementary to two distinct (and distant) locations on the scaffold. 
We expect that this design may cause the staples to ``bridge'' between different scaffold molecules. While this inter-scaffold bridging typically causes misfolding of DNA origami, we here consider this effect as part of the designability of the structures (see Fig.~\ref{fig:gelation}).

To test this hypothesis, we prepared dilute ($C \approx 0.1\ \mu g/\mu L < C^*$) and dense ($C \approx 1\ \mu g/\mu L > C^*$) solutions of linear and tadpole DNA origami polymers. Next, all four samples (two concentrations and two topologies) were heated to 70$^\circ$C for 15 minutes in a heat bath, and cooled to room temperature on the benchtop. By performing this thermal annealing above 60$^\circ$C, followed by quenching, we make sure that the DNA staples melt from the scaffold and re-hybridise, in turn re-folding the DNA origami polymer \emph{in situ}. Finally, aliquots were taken from each sample to run diagnostic gel electrophoresis and AFM. 

First, gel electrophoresis confirms what we expected, \emph{in situ} annealing and quenching of the linear DNA origami structure at both low and high concentrations does not affect the final folded structure. In other words, the design of the linear DNA origami ensures that the staples bind to the scaffold in the same way regardless of concentration, creating dsDNA which migrates the same distance as the 8 kbp dsDNA control band (see Figure \ref{fig:gelation}c, linear bands). On the contrary, \emph{in situ} refolding of the tadpoles leads to the formation of bands that ran slower than the control structure (see Figure \ref{fig:gelation}c, tadpole bands). We then extracted the DNA from the newly appeared band and visualised it in AFM, which revealed that, as expected, the refolded tadpoles formed a network of scaffolds (see Figure \ref{fig:gelation}b).
In other words, the majority of DNA tadpole structures were dismantled by the annealing process and replaced by a network of  scaffolds crosslinked by staples. 

We then conjectured that the formation of a network of DNA scaffolds should result in a significant slowing down of DNA dynamics and in the onset of gel-like behaviour. To test this hypothesis we performed  microrheology of each of the eight samples: linear and tadpole designs, refolded below/above $C^*$ and before/after \textit{in situ} refolding. 

First, we found that for both of the concentrations investigated (above and below $C^*$), the rheology of the linear topology remained unaffected by the \emph{in situ} refolding, and that the solution's viscosity remained unaffected ( Fig.~\ref{fig:gelation}d, left). This agrees with the structural observations in Fig.~\ref{fig:gelation}b-c. 

In marked contrast with the behaviour of the linear design, we observed that the rheology of the tadpole samples was affected by the \emph{in situ} refolding. More specifically, the MSD of the tracers in the solution of tadpole DNA origami below $C^*$ was mostly unaffected by the refolding (Fig.~\ref{fig:gelation}d). However, above $C^*$ we observed a dramatic shift of the MSD downward, reflecting a significant increase of the solution's viscosity (Fig.~\ref{fig:gelation}d). Specifically, we found that the viscosity of the tadpole solution above $C^*$ after the \emph{in situ} refolding increased $\approx$ 16 fold. Additionally, the MSDs displayed subdiffusive behaviour for the whole lagtime range recorded here (about two minutes) indicating constrained movement and in turn elastic, gel-like behaviour of the fluid at these timescales. 

\subsection{Reversible gelation via staple competition}
Having observed gelation by \emph{in situ} refolding, we then asked if we could leverage the equilibrium nature of DNA origami structures (and specifically strand displacement) to modify the solutions' rheology \emph{in situ} by adding competitor staples~\cite{Rossi2023Isothermal}. 

Opposite to isothermal reconfiguration of complex DNA origami structures at low concentrations as in Ref.~\cite{Rossi2023Isothermal}, here we want to investigate the rheological response of isothermal, \emph{in situ} reconfiguration of simple DNA origami structures at high concentrations, where the main complexity is found in how the scaffold interact with each other. 

Having demonstrated that re-folding in situ creates crosslinks between DNA scaffolds  mediated by looping staples, we now ask if by introducing linear staples that act as competitors to the looping staple, we could ``strand-displace'' the crosslinks and disassemble the network.

To test this conjecture, we created a crosslinked solution of tadpoles as in the previous section. The DNA origami folding solution contains 1:20 stoichiometry of scaffold (20 nM) to staple (400 nM). There are 4 staples in the linear staple stock that bind to the same regions as the 6 ``looping'' staples in the tadpole design. To create a competition, we therefore introduced a large excess of the 4 linear staples (final concentration 2 $\mu M$) in the tadpole sample. The sample was mixed thoroughly, and a dilution taken for gel electrophoresis analysis. The sample was then immediately loaded onto a slide for time-resolved microrheology measurements (2 minute videos of 2 fps were taken at 20 minute intervals, with `crosslinked tadpoles' marking the first video taken after the sample was loaded onto the slide). 

\begin{figure}[t!]
    \centering
    \includegraphics[width=1\linewidth]{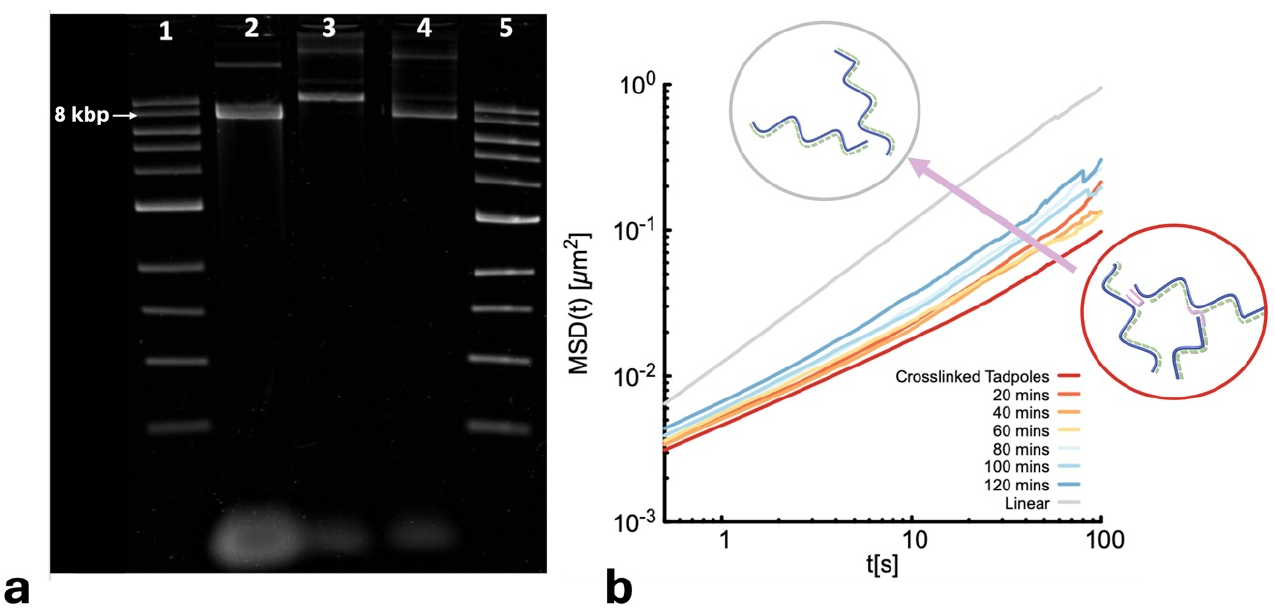}
    \caption{Gel electrophoresis showing reconfiguration of tadpole topology DNA origami to a linear topology. Linear DNA origami (lane 2), tadpole DNA origami (lane 3), tadpole DNA origami with excess linear staples in solution (lane 4). 1kbp ladder control (lanes 1 and 5) with 8 kbp labelled. Time-resolved microrheology of network disruption due to the presence of competitor staples (corresponding to the linear topology).}
    \label{fig:competitor}
\end{figure}

Figure \ref{fig:competitor}a shows gel electrophoresis of a linear origami control (lane 2), a tadpole control (lane 3) and a sample made by in situ refolding of tadpoles at $C>C^*$ where linear staples are added in excess (lane 4). The latter shows a band that travels at the same speed as the linear DNA structure, in turn suggesting that the strand displacement took place.

Motivated by these observations, we performed time-resolved microrheology where 2 minute-long movies were taken every 20 minutes after introducing the linear staples into the crosslinked tadpoles sample. 

As shown in Fig.~\ref{fig:competitor}b, the MSDs of the tracers increased over ``ageing'' time. This indicates that the competitor strands resolved the network of entanglements and the corresponding viscosity of the sample decreased $\approx$ 3-fold between the first data point (crosslinked tadpoles) and the final at 2 hrs (120 mins). We include the MSD of the linear alone for reference. Additionally, in this experiment, we were limited by the tracers' sedimentation which allowed us to track them only for 2 hours. 

Interestingly, in more conventional DNA origami systems, isothermal reconfiguration (Ref.~\cite{Rossi2023Isothermal}) occurred thanks to the large number of unpaired bases, which facilitated strand displacement. However, our tadpole DNA origami design does not contain any unpaired bases, and we still observe the effect of strand displacement and reconfiguration \emph{in situ}. We reasoned that this was likely due to a difference in size and geometric frustration between the staples. Indeed, the 4 linear staples are each 36 bp, whereas the 6 ``looping'' staples in the tadpole design are 23-26 bp in length, rendering them less stable~\cite{santalucia1998unified}. Additionally, the linear design staples do not stabilise a loop in the scaffold, and therefore are not subject to the free energy penalty due to looping entropy and electrostatic interactions of the charged DNA backbone experienced by the tadpole designs~\cite{Brackley2017Nonequilibrium}.


\section{Conclusions}

In summary, our work exemplifies how simple DNA origami structures can be used as a platform to build complex and ``topological'' polymers. 
Here we tried to manufacture large amounts of the simplest non-trivial topological (or chimeric) polymer, i.e. a tadpole, a polymer architecture featuring both circular and linear substructures. 
Motivated by recent simulations~\cite{Rosa2020} and experimental~\cite{Doi2020tadpole} work, we expected the tadpoles to display a significant, cooperative, slowing down in their dynamics due to threading~\cite{Michieletto2017prl,Michieletto2016pnas} (Fig.~\ref{fig:origamidesign}). While we provide strong evidence that we successfully designed linear, circular and tadpole DNA origami polymers (Fig.~\ref{fig:origamidesign}), unfortunately we did not observe any signature of the different architecture on the rheology of the samples (Fig.~\ref{fig:MSD_TLC_2fps}). Instead, we observed scaling laws suggesting that the DNA origami scaffolds are not long enough to display deep enough threadings to affect the dynamics of the polymers~\cite{Ubertini2022Entanglement}.

However, our findings that the viscosity and elasticity of the samples follow scaling laws expected in classic polymer theories confirm the idea that DNA origami structures can be used to make polymer-like objects that behave according to polymer physics (Fig.~\ref{fig:master}).  

Motivated by the dynamic and reconfigurable nature of DNA origami structures~\cite{Rossi2023Isothermal}, we explored the design of in situ refolding and design-dependent crosslinking of DNA origami scaffolds. In other words, we showed that fluids made of DNA origami can easily be made thermally responsive by careful design of the staple sequence. Specifically,  ``looping'' staples will bridge scaffolds and create physical crosslinks that can be tuned by staple sequence, length and position along the scaffold (Fig.~\ref{fig:gelation}). Interestingly, given that the networks are made by DNA-mediated crosslinks, we can leverage strand-displacement reactions to disassemble the network at room temperature simply by introducing competitor strands (Fig.~\ref{fig:competitor}).  

In the future, it would be interesting to design DNA origami polymers made by fusing DNA origami scaffolds. In this way, we could extend structures to 16 kbp (two scaffolds) or 32 kbp (three scaffolds). However, we expect the yield to be smaller which raises technical limitations with performing rheology at high concentrations. 
An alternative solution to increase the likelihood of threading would be to make the DNA structures more rigid by phosphorylating the 5' ends of the staples and using T7 ligase to covalently close the double helical backbone along the origami. In this way the structures would be made of dsDNA which is significantly more rigid than nicked DNA.

\section{Acknowledgements}
DM acknowledges the Royal Society and the European Research Council (grant agreement No 947918, TAP) for funding. The authors also acknowledge the contribution of the COST Action Eutopia, CA17139 and Soft Matter for Formulation and Industrial Innovation (SOFI CDT) (grant reference EP/S023631/1). For the design of the DNA origami tadpoles, we acknowledge technical assistance from Tilibit Nanosystems and in particular Tamara Aigner and Jean-Philippe Sobczak. For acquiring the AFM images,  we acknowledge technical help from Kislon Voitchovsky (University of Durham), Allard Katan (TU Delft), Laura Charlton (Edinburgh), Alice Pyne and Libby Holmes (University of Sheffield). We wish to acknowledge the support of the Henry Royce Institute for advanced materials and the Student Equipment Access Scheme which enabled access to Bruker Dimension XR facilities at The Royce Discovery Centre at the University of Sheffield; EPSRC Grant Number EP/R00661X/1. We acknowledge Patricia Gonzalez Iglesias and Tracy Scott for their technical support and assistance in the laboratory, and Jochen Arlt for microscopy training and assitance for microrheology. For the purpose of open access, the author has applied a Creative Commons Attribution (CC BY) licence to any Author Accepted Manuscript version arising from this submission. 

\section{Conflicts of Interest}
There are no conflicts of interest to declare.

\bibliography{biblio}
\end{document}